\documentclass[%
aip,
 amsmath,amssymb,
reprint,%
]{revtex4-1}

\usepackage{multirow}
\usepackage{graphicx}
\usepackage{bm}% bold math
\usepackage{hyperref}% add hypertext capabilities
\usepackage[margin=.7in]{geometry}

\usepackage[version=3]{mhchem} 
\usepackage{chemarr}

\usepackage{mathrsfs}

\begin{document}

\title{Information Encoding/Decoding using the Memory Effect in Fractional-order Capacitive Devices}  

\author{Anis Allagui}
\email{aallagui@sharjah.ac.ae}
\affiliation{Dept. of Sustainable and Renewable Energy Engineering, University of Sharjah, PO Box 27272, Sharjah, United Arab Emirates}
\affiliation{Center for Advanced Materials Research, Research Institute of Sciences and Engineering, University of Sharjah, PO Box 27272, Sharjah, United Arab Emirates}
\altaffiliation[Also at ]{Dept. of Mechanical and Materials Engineering, Florida International University, Miami, FL33174, United States}

\author{Ahmed S. Elwakil}
\affiliation{Dept. of Electrical and Computer Engineering, University of Sharjah, PO Box 27272, Sharjah, United Arab Emirates}
\altaffiliation[Also at ]{Nanoelectronics Integrated Systems Center (NISC), Nile University, Cairo, Egypt}
\altaffiliation[and ]{Dept. of Electrical and Computer Engineering, University of Calgary, Calgary, Canada}
%
%
%
%\author{Khaled N. Salama}
%\affiliation{King Abdullah University of Science and Technology (KAUST), Saudi Arabia}
%

\begin{abstract}

%In this study we provide an experimental proof of the existence of a memory effect in supercapacitors.  In particular, 

In this study, we show that the discharge voltage  pattern of a fractional-order supercapacitor from the same initial steady-state voltage into a constant resistor is dependent on  the past charging voltage profile. The charging voltage was designed to follow a power-law function, i.e. $v_c(t)=V_{cc} \left( {t}/{t_{ss}}\right)^p \;(0<t \leqslant t_{ss})$, in which $t_{ss}$ (charging time duration between zero voltage to the terminal voltage $V_{cc}$) and $p$ ($0<p<1$) act as two variable parameters.  
We used this history-dependence of the dynamic behavior of the device to uniquely  retrieve     information pre-coded in the  charging waveform pattern. Furthermore, we provide an analytical model based on fractional calculus that explains phenomenologically the information storage mechanism.   The use of this intrinsic material memory effect may lead to   new types of  methods for information storage and retrieval. 

\vspace{1cm}

\end{abstract}

\maketitle

\section*{Introduction}

\begin{figure}[h]
\begin{center}
\includegraphics[width=0.35\textwidth]{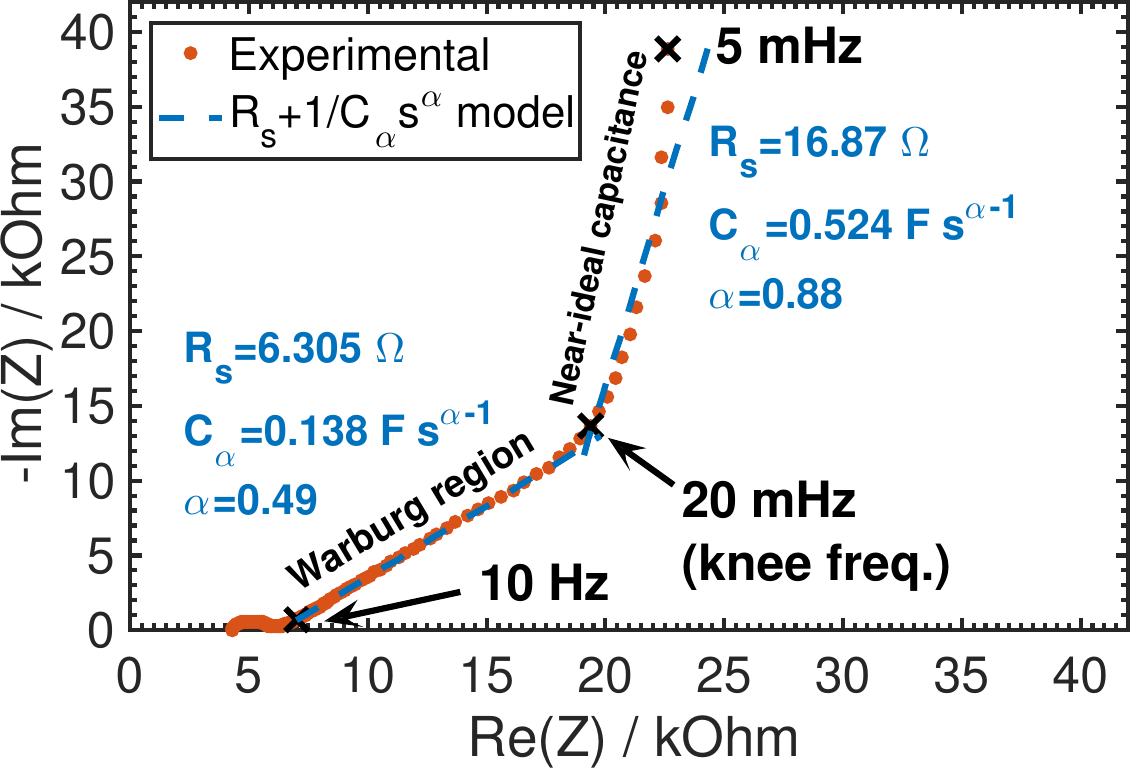}
\caption{Nyquist plane representation  of open-circuit spectral impedance  of a NEC/TOKIN supercapacitor (part \#FGR0H105ZF, 5.5\,V, 1\,F). Complex nonlinear least-squares fitting to $Z(s)=R_s+1/C_{\alpha}s^{\alpha}$ ($s=\mathrm{j}\omega$) shows two straight line regions giving the values $(R_s;C_{\alpha};\alpha)=(6.306\;\Omega;\;0.138\;\text{F\,s}^{\alpha-1};\;0.49)$ from    10 Hz to 20 mHz, and $(16.87\;\Omega;\;0.524\;\text{F\,s}^{\alpha-1};\;0.88)$ from 20 mHz to 5 mHz.  20\,mHz is a critical frequency separating the near-ideal capacitive behavior from the Warburg region}
\label{EIS}
\end{center}
\end{figure}

Many systems and processes in nature exhibit fractional-order behavior, such as 
vestibulo-ocular reflex \cite{anastasio1994fractional}, 
neuronal activity \cite{teka2014neuronal},  
ion channel gating \cite{goychuk2004fractional}, 
viscoelasticity \cite{mainardi2011creep},  and  
disordered semiconductors \cite{scher1975anomalous},  which are models known to capture the existing short/long-term memory effects in these systems \cite{sabatier2014long, teka2017fractional, yuan2014extracting, ventosa2014long}. 
Supercapacitors  are well-established types of electrochemical capacitive energy storage devices, but are also known to 
 exhibit non-ideal, time-fractional-order electric behavior when charged by a power supply and  discharged into a load \cite{kant_generalization_2015, eis,2018-1, energy2015, 2018-2, 2017-3}. 
Their electric impedance can be modeled  as a series resistance ($R_s$) with a fractional-order capacitor (also known as constant-phase element, CPE) over a certain frequency range; the reduced impedance of such a model is   
$Z^*(u)=1+ {1}/{(\mathrm{j} u)^{\alpha}}$ 
with $u=\omega (R_sC_{\alpha})^{1/\alpha}$,
$C_{\alpha}$ is a fractional-order capacitance in units of F\,s$^{\alpha-1}$, 
 $\alpha$ ($0<\alpha < 1$) is a dimensionless fractional exponent, 
 and
 $\omega$ is the applied angular frequency in units of s$^{-1}$ (see Nyquist plot of a commercially-available NEC/TOKIN supercapacitor in Fig.\;\ref{EIS}). 
In the time-domain, the current-voltage relationship of an $R_s$-CPE-equivalent supercapacitor 
  is expressed by the fractional-order differential equation \cite{ EC2015, fracorderreview}: 
 
\begin{equation}
   R_s C_{\alpha} \frac{d^{\alpha}V_{C_{\alpha}}(t)}{dt^{\alpha}} +  V_{C_{\alpha}}(t)  = v_c(t)
\label{eq:2}
\end{equation}
where $V_{C_{\alpha}}$  is the voltage across the CPE part ($V_{C_{\alpha}}=0$ for $t\leq0$), 
 $C_{\alpha} {d^{\alpha}V_{C_{\alpha}}(t)}/{dt^{\alpha}}$ is the current flowing through the CPE, and $v_c(t)$ is the applied charging voltage. 
The differentiation of   non-integer order in equation\;\ref{eq:2} is defined as \cite{podlubny1998fractional}: 

\begin{equation}
\frac{d^{-\alpha} {V}(t)}{dt^{-\alpha}} = \frac{1}{\Gamma(\alpha)} \int\limits_0^t   {V}(\tau) (t-\tau)^{\alpha-1} d\tau , 
\label{eq:1} 
\end{equation}
which  
can be viewed as a convolution with a hyperbolic function of frequency, and therefore contains a memory that progressively increases as the fractional order $\alpha$ decreases \cite{westerlund1994capacitor, westerlund1991dead, du2013measuring}. 
 The physical interpretation of such fractional-order electric behavior of supercapacitors   is still under debate, nonetheless, it has been widely attributed to the surface chemistry and morphological structure of the electrodes, which are usually composed of high-surface area and porous materials separated by an ionic conductor \cite{doi:10.1021/jp512063f, alexander2015contribution, alexander2016contribution, alexander2017contribution}. 

Despite the fact that fractional-order models involve hereditary effects, very few robust experimental proofs of this memory have been reported and investigated  \cite{westerlund1991dead, westerlund1994capacitor, uchaikin2016memory, uchaikin2009memory, memoryAPL, memQ}. 
This Letter illustrates the memory effect in supercapacitors by  sequentially encoding  information into the charging pattern of the device, and then uniquely retrieving this code  from the discharge pattern. 
The purpose here is to investigate this effect experimentally and  mathematically using fractional-order models, which means that we are leaving the physical interpretation and potential applications of this effect to another study.
%Subsequently, the information can be uniquely  decoded from the discharge pattern. 
The results are obtained on state-of-the-art, commercially-available devices which offer excellent  stability, and relatively high-voltage ratings which facilitate repeatability and precise measurements.

\section*{Methods}

We applied different charging voltage waveforms to a NEC/TOKIN supercapacitor (part \#FGR0H105ZF, 5.5\,V, 1\,F\footnote{This device consists of  six 0.917\,V-aqueous electrolytic cells stacked in series; each cell is composed of two symmetric activated carbon + dilute sulfuric acid electrodes separated by a porous organic film}) following the power-law function: 

\begin{equation}
v_c(t)=V_{cc} \left(\frac{t}{t_{ss}}\right)^p \quad (0<t \leqslant t_{ss})
\label{eq1}
\end{equation}
where $p$ is an exponent taking values between 0 (step voltage) and 1 (linear ramp), and $t_{ss}$ is the rise time from 0\,V to the steady-state value $V_{cc}$ ($V_{cc}=5.5\;\text{V}$ for this device), after which the  charging voltage  is turned off. The rise time  $t_{ss}$ is pre-defined  so that the device will operate either in the capacitive tail from 20\,mHz to 5\,mHz, or in the Warburg region from 10\,Hz to 20\,mHz (see the two quasi-linear regions in  Fig.\;\ref{EIS}). 
Prior to each applied charging voltage waveform, the supercapacitor was fully discharged into a constant 100\;$\Omega$ resistor ($R_p$) until its voltage was equal to 1\,mV. 
Depending on the values of $p$ and $t_{ss}$, the charge waveforms used for information storage and the discharge waveforms used for information retrieval are given by the letter/number codes in Table \ref{tab1}.
All charging/discharging experiments of the supercapacitor  were programmed and executed sequentially   on a  Bio-logic VSP-300 electrochemical station using the EC-Lab control software.
The time step for collecting data in all  measurements was set to 10\,ms.

\begin{table}[h]
\caption{Table of codes illustrating encoding and decoding waveforms symbols in the form of different values of $p$ and $t_{ss}$ in $v_c(t)=V_{cc} ({t}/{t_{ss}})^p$. {We will be using the superscripts $"c"$ and $"d"$ for charge and   discharge, respectively. For example $A^c_{10}$} refers to the code $A_{10}$ during charge while $A^d_{10}$ refers to the same code during discharge.}
\begin{center}
\begin{tabular}{l | lllll}
{$p$}  \textbackslash {$t_{ss}$} & 550\,s & 275\,s & 110\,s & 55\,s & 27\,s  \\[1ex] \hline
 1.0 & $A _{10}$ & $B _{10}$ &  $C _{10}$ & $D _{10}$   & $E _{10}$  \\[1ex]
 0.7 & $A _{07}$ & $B _{07}$ &  $C _{07}$ & $D _{07}$   & $E _{07}$  \\[1ex]
 0.4  & $A _{04}$ & $B _{04}$ &  $C _{04}$ & $D _{04}$   & $E _{04}$  \\[1ex]
  0.2  & $A _{02}$ & $B _{02}$ &  $C _{02}$ & $D _{02}$   & $E _{02}$  \\[1ex]
   0.1 & $A _{01}$ & $B _{01}$ &  $C _{01}$ & $D _{01}$   & $E _{01}$ 
\end{tabular}
\end{center}
\label{tab1}
\end{table}%

\section*{Results}

\begin{figure}[h]
\begin{center}
\includegraphics[width=0.4\textwidth]{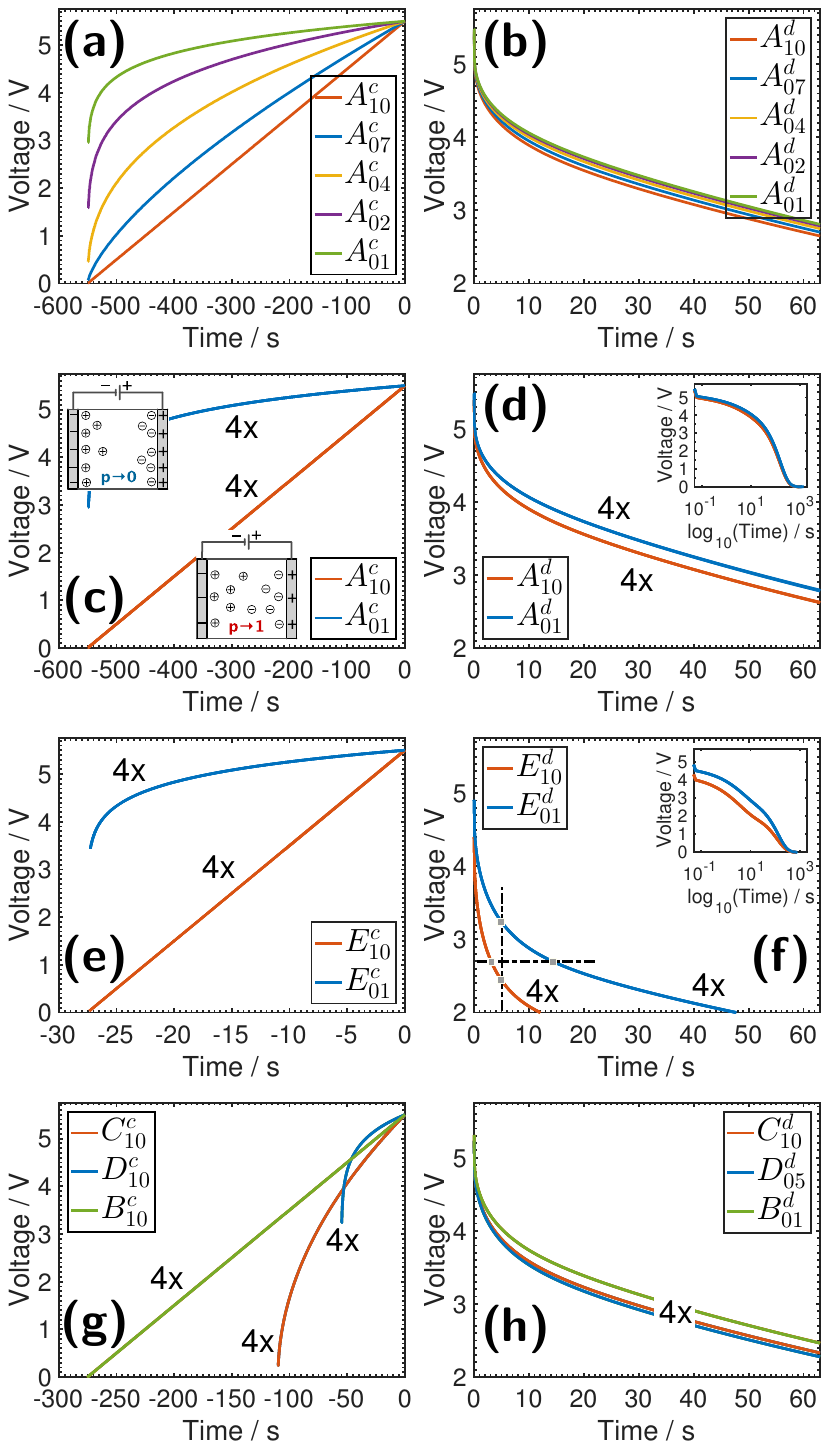}
\caption{Charging sequence (left) using a power supply with a voltage $v_c(t)=V_{cc} \left({t}/{t_{ss}}\right)^{p}$,  and discharging sequence  (right)  into a constant 100\,$\Omega$ resistor of  an NEC/TOKIN supercapacitor. (a)-(b) depict the  charge/discharge voltage patterns  for different values of $p$ (1.0, 0.7, 0.4, 0.2 and 0.1) and $t_{ss}=550$\,s (table\;\ref{tab1}), i.e. $A^c_{10}$/$A^d_{10}$-$A^c_{07}$/$A^d_{07}$-$A^c_{04}$/$A^d_{04}$-$A^c_{02}$/$A^d_{02}$-$A^c_{01}$/$A^d_{01}$. (c)-(d)  and  (e)-(f) show the repeatability of the process for four consecutive cycles (4x, superimposed on top of each other) of charge/discharge with $p=1.0$ and $p=0.1$; in (c)-(d)  waveforms $A^c_{10}$/$A^d_{01}$-$A^c_{10}$/$A^d_{01}$-$A^c_{10}$/$A^d_{01}$-$A^c_{10}$/$A^d_{01}$, and in (e)-(f)   waveforms $E^c_{10}$/$E^d_{01}$-$E^c_{10}$/$E^d_{01}$-$E^c_{10}$/$E^d_{01}$-$E^c_{10}$/$E^d_{01}$. In (g)-(h), we applied the sequence of charge/discharge codes 
[$B^c_{10}$/$B^d_{10}$-
$C^c_{05}$/$C^d_{05}$-
$D^c_{01}$/$D^d_{01}$]--[$B^c_{10}$/$B^d_{10}$-
$C^c_{05}$/$C^d_{05}$-
$D^c_{01}$/$D^d_{01}$]--[$C^c_{05}$/$C^d_{05}$-
$D^c_{01}$/$D^d_{01}$-
$B^c_{10}$/$B^d_{10}$]--[$C^c_{05}$/$C^d_{05}$-
$D^c_{01}$/$D^d_{01}$-
$B^c_{10}$/$B^d_{10}$]
 using three  values of $t_{ss}$ and three  values of $p$ (Table\;\ref{tab1}) demonstrating the possibility of two-dimensional encoding as well as the repeatability of the process   (4x)}
\label{cd}
\end{center}
\end{figure}

The experimental results are shown in Fig.\;\ref{cd}. In the first column of the figure, we show several combinations of the charging voltage waveforms applied to the supercapacitor by varying the values of $p$ and $t_{ss}$ (see Table\;\ref{tab1}).   The time scale is shifted by $-t_{ss}$ to represent past events.
 The second column of the figure shows the first 60\,s of the resulting voltage discharge waveforms into the same  100\;$\Omega$ resistor. {Here the potentiostat acts as a   constant resistance  by controlling the current to maintain the ratio voltage/current constant.} 
  Fig.\;\ref{cd}(b) shows five different voltage discharge profiles ($A_{10}^d$, $A_{07}^d$, $A_{04}^d$, $A_{02}^d$, $A_{01}^d$) after the device was charged with five different voltage waveforms ($A_{10}^c$, $A_{07}^c$, $A_{04}^c$, $A_{02}^c$, $A_{01}^c$)  (Fig.\;\ref{cd}(a)). The value of $t_{ss}$ is made long enough so that the device  will operate within its low frequency capacitive tail (Fig.\;\ref{EIS}). All discharging waveforms  show first a quick voltage drop of \emph{ca.} 0.5\,V from the initial voltage $V_{cc}=5.5\;\text{V}$ into the internal series resistance of the device ($R_s$), followed by a non-Debye inverse power law profile (i.e. $v_d(t) \propto t^{-\alpha}$  \cite{jonscher1977universal, westerlund1994capacitor, westerlund1991dead, jonscher1999dielectric}). 
  In Fig.\;\ref{cd}(d), we show the discharge profiles of the device for four consecutive cycles (4x) alternating  between code $A_{10}^c$ and code $A_{01}^c$ (Fig.\;\ref{cd}(c)). 
  The corresponding discharge waveform codes $A_{10}^d$ and $A_{01}^d$ are perfectly superimposed on each other demonstrating the repeatability of the charge/discharge process and the good stability of the device between one sequence to another. The results in Figs\;\ref{cd}(a)-(d) show that the supercapacitor voltage discharge profile into the same constant resistance depends on the exponent $p$ in the charging voltage profile, and thus on its prehistory. This would not be the case for an ideal capacitor for which the exponential decaying  voltage  $v_d(t) = V_{cc}  \exp  \left(-{t}/{R_pC}\right)$ depends on the initial voltage $V_{cc}$, but not on how this voltage has been reached..
 {In other words, information can be encoded in the exponent $p$ being  for example $1.0$ or $0.1$ (Figs.\;\ref{cd}(c) and \ref{cd}(d)) which can be used to encode a binary data sequence, or in multi-level logic using multiple values of $p$  (Figs.\;\ref{cd}(a) and \ref{cd}(b))}.

The effect of the parameter $t_{ss}$ is examined in   Figs.\;\ref{cd}(e) and \;\ref{cd}(f), which depict the voltage profiles that were obtained in a similar way to the results shown in Figs.\;\ref{cd}(c) and \;\ref{cd}(d), respectively, but with a faster charging rate ($t_{ss}=27\,\text{s}$). 
 This value of $t_{ss}$ corresponds to about 37\;mHz frequency which  belongs to the Warburg region and not the capacitive tail anymore (see Fig.\;\ref{EIS}). It is clear from Fig.\;\ref{cd}(f) that the superposition of the code plots ($E_{10}^d$ and $E_{01}^d$) obtained in response to the  charging voltage codes $E_{10}^c$ and $E_{01}^c$ is again impeccable. Additionally,  the difference between the discharge waveforms   is more pronounced than when $t_{ss}$ was set to $550\,\text{s}$ (Fig.\;\ref{cd}(d)) at which point the supercapacitor behaved as a near-ideal capacitor.  Although the values of $R_s$ and $C_{\alpha}$ are not same for these two cases,  the correlation between the discharging and the charging voltage waveforms is stronger as $\alpha$ decreases (equations\;\ref{eq:2} and \ref{eq:1}) which makes $t_{ss}$ another possible information coding dimension. 
 
{As for the decoding of the discharge pattern corresponding to a specific charge pattern}, this can be carried out by a simple adaptive thresholding method either for a fixed time or fixed voltage. 
  For example, as shown in Fig.\;\ref{cd}(f),  the thresholding can be performed at 5\,seconds from the beginning of the discharge giving the two voltage values of 2.441 and 3.238\;V  for $p=1.0$ and $p=0.1$ (i.e. a difference of 0.797\;V), respectively. Alternatively, it can be performed at a fixed voltage value of 2.7\,V for example, leading to decoded time intervals of 3.109 and 14.31\;s for $p=1.0$ and $p=0.1$, respectively. %It is important to note that as the device approaches its steady-state, the difference between the voltage discharge patterns decays . 
It is important to note that decoding from the discharging waveform should be carried out way before the device is fully discharged, as shown in decimal algorithm scale in the  insets in Figs.\;\ref{cd}(d) and \;\ref{cd}(f) depicting  the quick convergence of the voltage-time profiles. The reason is that fractional-order behavior is  space- and time-dependent; i.e. only in the transient time does the order of the state-space (represented by the values of $\alpha$) manifest itself leading to the memory effect \cite{memoryAPL,memQ}. In the steady-state time, fractional-order behavior becomes space-order independent.
 
Another set of experimental results are shown in Figs.\;\ref{cd}(g) and\;\ref{cd}(h), in which charge/discharge of the device followed  the sequence [$B_{10}$-$C_{05}$-$D_{01}$]--[$B_{10}$-$C_{05}$-$D_{01}$]--[$C_{05}$-$D_{01}$-$B_{10}$]--[$C_{05}$-$D_{01}$-$B_{10}$]. {This indicates the possibility of  two-dimensional memory encoding using $p$ and $t_{ss}$}.  

%\begin{figure}[t]
%\begin{center}
%\includegraphics[width=0.49\textwidth]{fig3.eps}
%\caption{Information storage (write) and retrieval (read) of an 8-bit binary ASCII code 00011010 using $C_{10}$ for "0" and $C_{01}$ for "1".}
%\label{ascii}
%\end{center}
%\end{figure}
% 
%{Finally, in Fig.\;\ref{ascii} we show eight successive  charge/discharge cycles into the supercapacitor corresponding to the digital code 00011010. The digital logic "0" is encoded as $C_{10}$, while the digital logic "1" is encoded as $C_{01}$. The information retrieval and its repeatability are clear. The same process can be extended to multi-logic encoded data in two dimensions as illustrated in table\;\ref{tab1} if both $t_{ss}$ and $p$ are taken together as variable parameters in the charging waveform.}

\section*{Discussion}

Analytically, the memory effect can be explained as follows. The voltage discharge response of the supercapacitor, modeled as an $R_s$-CPE equivalent circuit, into a parallel resistance $R_p$ %after being charged with $v_c(t) = V_{cc} \left({t}/{t_{ss}}\right)^p$, 
is given by \cite{freeborn2013measurement}:
\begin{equation}
v_d(t) = v_d(0) \frac{R_p}{R_p+R_s} \text{E}_{\alpha,1}\left(-  \frac{t^{\alpha}}{\tau_p^{\alpha}+\tau_s^{\alpha}}  \right)
\label{eq6}
\end{equation}
where $\tau_p=\left( R_p C_{\alpha} \right)^{1/\alpha}$, $\tau_s=\left( R_s C_{\alpha} \right)^{1/\alpha}$,  $\text{E}_{\alpha,\beta}(z)= \sum_{k=0}^\infty z^k/\Gamma(\alpha\,k+\beta)$ with ($\alpha>0$, $\beta>0$) is the two-parameter Mittag-Leffler function, and the initial voltage $v_d(0)$ is given by
$v_d(0) = V_{cc} - R_s i_c(t_{ss})$ where $i_c(t)$ is the charging current obtained as $dq_c(t)/dt$ (see equation \ref{eq9}). For $R_p \gg R_s$, 
equation\;\ref{eq6} can be rewritten in a dimensionless form as follows:

\begin{equation}
\tilde{v}_d{(\tilde{t}_p)} \simeq R_{(p,m,\alpha)} \text{E}_{\alpha,1} \left(- \tilde{t}_p^{\,\alpha} \right)
\label{eq7}
\end{equation}
where $\tilde{v}_d{(\tilde{t}_p)}=v_d(t)/V_{cc}$, 
$\tilde{t}_p =(t/\tau_p)$ 
and $m=(t_{ss}/\tau_s)$. The Mittag-Leffler function in equation\;\ref{eq7}  depends   on the CPE parameters $\alpha$ and $C_{\alpha}$ (and on $R_p$), whereas the parametric function $R_{(p,m,\alpha)}$ is dependent also on   how the device has been charged via the selection of one or both of the applied waveform parameters $p$ and $t_{ss}$. For an ideal capacitor, i.e. with $\alpha=1$ and $R_s=0$, we verify that equation\;\ref{eq7} simplifies to $ \tilde{v}_d{(\tilde{t}_p)} = \exp(- \tilde{t}_p)$ which does not depend on its charge prehistory, as expected.

\begin{figure}[t]
\begin{center}
\includegraphics[width=0.45\textwidth]{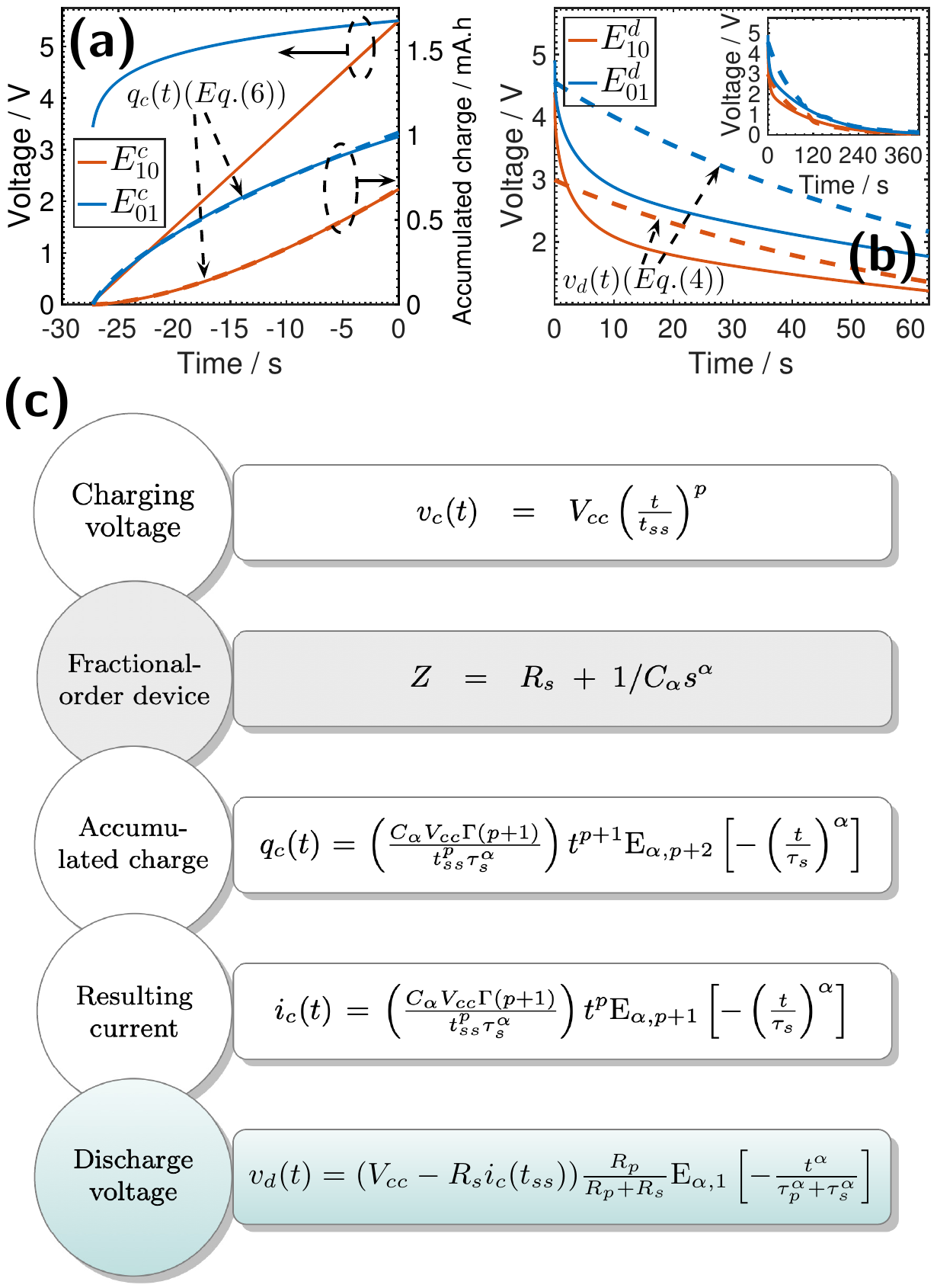}
%\begin{tikzpicture}[font=\sffamily\large]
%\draw (-0.85in,0) node[inner sep=0]{\includegraphics[height=1.4in]{NEC-088-095-028s-1p0-0p1-PPI-vt-CLD-vt-sim2.eps}}; % [a0 Ca0 Rs0] = [0.4797    0.0340 6.3051]
%\draw[black] (-1.35cm-0.85in, 1.42cm) node {\textbf{(a)}};
%%\draw[black] (-0.1cm, 2.1cm) node {\textbf{Information storage}};
%\draw (.95in,0) node[inner sep=0]{\includegraphics[height=1.4in]{NEC-088-095-028s-1p0-0p1-CLD-vt-sim2.eps}}; % [a0 Ca0 Rs0] = [0.8647    0.5998  6.305]
%\draw[black] (-1.48cm+3.02in-0.85in, -0.65cm) node {\textbf{(b)}};
%%\draw[black] (-0.175cm+1.9in, 2.13cm) node {\textbf{Information retrieval}};
%\draw (0,-2.5in) node[inner sep=0] {\includegraphics[width=0.47\textwidth]{diagram/diagram}};
%\draw[black] (-1.35cm-1.1in, 1.42cm-1.37in) node {\textbf{(c)}};
%\end{tikzpicture}
\caption{
(a) Supercapacitor charging using the voltage waveforms $v_c(t)=V_{cc} \left({t}/{t_{ss}}\right)^{p}$ with $t_{ss}=27\;\text{s}$ and $p=1.0$ and $0.1$ (i.e. $E_{10}^c$ and $E_{01}^c$) and the resulting time-charge profiles. 
(b) Illustrates the corresponding discharging voltage waveforms (i.e. $E_{10}^d$ and $E_{01}^d$). Experimental measurements are plotted in solid lines and simulations  are represented by  dashed lines. 
(c)   Flowchart representing the process and equations of the electric variables during charge (i.e. applied voltage, accumulated charge, and corresponding current), and voltage during discharge  into a constant resistor (the device is considered. to be a fractional-order supercapacitor of impedance $Z=R_s+1/C_{\alpha}s^{\alpha}$ and  being charged with a voltage function $v_c(t)=V_{cc} \left({t}/{t_{ss}}\right)^p$)}
\label{fig3}
\end{center}
\end{figure} 

Now, to show the memory relationship, we derive the expression for the electrical charge stored in the device   subjected to the charging waveform $v_c(t) = V_{cc} \left({t}/{t_{ss}}\right)^p$  as follows \cite{memoryAPL}: 
% [{Equation\;\ref{eq8} is obtained by equating $i(t) = dq/dt$ with $C_{\alpha} {d^{\alpha}V_{C_{\alpha}}(t)}/{dt^{\alpha}}$ where $V_{C_{\alpha}}(t) $ is the solution of the fractional-order differential equation $ V_{cc} \left({t}/{t_{ss}}\right)^p = V_{C_{\alpha}} + R_s C_{\alpha} {d^{\alpha}V_{C_{\alpha}}(t)}/{dt^{\alpha}}$ obtained using the inverse Laplace transform identity $\mathscr{L}^{-1} \left(  \frac{k! s^{\alpha-\beta}}{(s^{\alpha} + \lambda )^{k+1} }    \right) = t^{\alpha\,k +\beta-1} E_{\alpha,\beta}^{(k)} [ - \lambda t^{\alpha} ] $ with ($k=0$, $\beta=p+1$, $\lambda=\tau_s^{-\alpha}$) \cite{podlubny1998fractional}.}]:

\begin{equation}
q_c(t) = \left( \frac{C_{\alpha} V_{cc} \Gamma(p+1)}{ t_{ss}^p \tau_s^{\alpha} } \right)  t^{p+1} \text{E}_{\alpha,p+2} \left[ - \left(  \frac{t}{\tau_s} \right)^{\alpha} \right] \vspace{.1cm}
\label{eq8}
\end{equation}
from which the current is found to be:
\begin{equation}
i_c(t) = \left( \frac{C_{\alpha} V_{cc} \Gamma(p+1)}{ t_{ss}^p \tau_s^{\alpha} } \right)  t^{p} \text{E}_{\alpha,p+1} \left[ - \left(  \frac{t}{\tau_s} \right)^{\alpha} \right] \vspace{.1cm}
\label{eq9}
\end{equation}
[Note that {Equation\;\ref{eq8} is obtained by equating $i(t) = dq/dt$ with $C_{\alpha} {d^{\alpha}V_{C_{\alpha}}(t)}/{dt^{\alpha}}$ where $V_{C_{\alpha}}(t) $ is the solution of the fractional-order differential equation $ V_{cc} \left({t}/{t_{ss}}\right)^p = V_{C_{\alpha}} + R_s C_{\alpha} {d^{\alpha}V_{C_{\alpha}}(t)}/{dt^{\alpha}}$ obtained using the inverse Laplace transform identity $\mathscr{L}^{-1} \left(  \frac{k! s^{\alpha-\beta}}{(s^{\alpha} + \lambda )^{k+1} }    \right) = t^{\alpha\,k +\beta-1} E_{\alpha,\beta}^{(k)} [ - \lambda t^{\alpha} ] $ with ($k=0$, $\beta=p+1$, $\lambda=\tau_s^{-\alpha}$) \cite{podlubny1998fractional}}].  At $t=t_{ss}$, the steady-state charge $q_c(t_{ss})$ is a %given by:
%\begin{equation}
%Q_c(t_{ss}) = \left( \frac{C_{\alpha} V_{cc} \Gamma(p+1)}{ \tau_s^{\alpha} } \right)  t_{ss} \; \text{E}_{\alpha,p+2} \left[ \left( \frac{t_{ss}}{\tau_s} \right)^{\alpha} \right]
%\end{equation}
 function  of both  parameters of the charge waveform ($p$ and $t_{ss}$) and those of the supercapacitor. [{Equation \ref{eq8} simplifies to $q_c(t) = CV_{cc}t/t_{ss}$ for an ideal capacitance ($C_{\alpha}=C$ and $\alpha=1$) using the identity $E_{1,3}(z) = (\exp(z)-z-1)/z^2$ which applies when $p=1$. If $p = 0$ (step voltage), the charge $q_c(t)=CV_{cc}$, otherwise $q_c(t)$ is a function of $p$}]. This is in line with our recent findings in which we highlighted that charging a supercapacitor with a voltage input results in a device and waveform-dependent accumulated electric charge \cite{2018-1,memoryAPL,memQ}. In a dimensionless form, equation\;\ref{eq8}   looks like this:

\begin{equation}
\tilde{q}_c(\tilde{t}_s)= S_{(p,m,\alpha)} \tilde{t}_s^{\,p+1} \text{E}_{\alpha,p+2} \left( -\tilde{t}_s^{\,\alpha} \right)
\end{equation}
where $\tilde{t}_s=t/\tau_s$. 
In Fig.\;\ref{fig3}(a), we show the measured and simulated (using equation\;\ref{eq8}) charge function $q_c(t)$ for a fixed value of $t_{ss}$ equal to 27\;s, and steady-value voltage of 5.5\,V.  Two values of the parameter $p$ (i.e. 1.0 and 0.1) are selected. Equation\;\ref{eq8} is in excellent agreement with the experiment using the values  
($R_s, C_{\alpha}, \alpha$) equal to 
($10\,\Omega,  35\,\text{mF\,s}^{\alpha-1}, 0.48$) 
for $p=1.0$, and  equal to 
($10\,\Omega, 36\,\text{mF\,s}^{\alpha-1}, 0.49$) 
for $p=0.1$ obtained using least-squares fitting.  
Experimentally, the accumulated charge at time $t_{ss}$ is found to be 0.68 and 0.99\,mA\,h$^{-1}$ for $p=1.0$ and $0.1$, respectively. {Given that all parameters in equation\;\ref{eq8} are fixed apart from $p$, it is evident that the information  encoded in the value of  $p$ is stored as $q_c(t)$.} 
In Fig.\;\ref{fig3}(b), we show the resulting discharge voltage along with simulation using equation\;\ref{eq6}. The best fit is found using the values 
($R_s, C_{\alpha}, \alpha$) equal to 
($10\,\Omega,  803\,\text{mF\,s}^{\alpha-1}, 0.96$) 
for $p=1.0$, and equal to 
($14.24\,\Omega,  855\,\text{mF\,s}^{\alpha-1}, 0.95$) 
for $p=0.1$, which can be improved if a sliding mode fitting is adopted instead.
Note that the increase of the values of $\alpha$ and $C_{\alpha}$ towards those of a near-ideal capacitance is the result of the slow discharge rate.

\section*{Conclusion}

%\begin{figure}[t]
%\begin{center}
%\includegraphics[width=0.49\textwidth]{diagram/diagram}
%%\begin{tikzpicture}[font=\sffamily\large]
%%\draw (0,0) node[inner sep=0]{\includegraphics[height=1.4in]{NEC-088-095-028s-1p0-0p1-PPI-vt-CLD-vt-sim2.eps}}; % [a0 Ca0 Rs0] = [0.4797    0.0340 6.3051]
%%\draw[black] (-1.35cm, 1.42cm) node {\textbf{(a)}};
%%\draw[black] (-0.1cm, 2.1cm) node {\textbf{Information storage}};
%%\draw (1.8in,0) node[inner sep=0]{\includegraphics[height=1.4in]{NEC-088-095-028s-1p0-0p1-CLD-vt-sim2.eps}}; % [a0 Ca0 Rs0] = [0.8647    0.5998  6.305]
%%\draw[black] (-1.48cm+3.02in, -0.65cm) node {\textbf{(b)}};
%%\draw[black] (-0.175cm+1.9in, 2.13cm) node {\textbf{Information retrieval}};
%%\end{tikzpicture}
%\caption{Electric parameters (charge, current and voltage) resulting from the discharge of a fractional-order supercapacitor of impedance $Z=R_s+1/C_{\alpha}s^{\alpha}$ into a constant resistor after being charged with a voltage function $v_c(t)=V_{cc} \left({t}/{t_{ss}}\right)^p$}
%\label{fig5}
%\end{center}
%\end{figure}

We have shown that the voltage discharge of a supercapacitor that exhibits fractional-order temporal dynamics  depends uniquely on the way by which it was charged. Figure\;\ref{fig3}(c)  recapitulates the process from  charging the device by an external power supply, to  the accumulation of charge and the creation of current, and their relationship with the discharge voltage function which depends on the voltage charging parameters $p$ and $t_{ss}$. In other words, the supercapacitor remembers the pattern by which it was charged, and as a result, discharges accordingly.
%Hence, information can be encoded in the voltage charge profile of such devices, and can be subsequently decoded from the discharge profile. 
%The use of supercapacitors for this purpose or \emph{materials} with one or more of their state variables exhibiting fractional-order behavior,  can be viewed as an analog memory with no  dc voltage source  required. 

The experimental results reported here were obtained on a commercial, high-capacitance supercapacitor   which resulted in   slow information storage and retrieval. 
%While this guarantees the repeatability of our experiments due to the excellent stability of these devices, 
Higher read-write rates should be possible in principle by using low-capacitance,  fractional-order devices \cite{CELC:CELC201700663}. However, it is important in this case to ensure  stability over a large number of cycles,  as well as to reduce the effect of any parasitic capacitance for properly assessing the memory effect.

%\section*{Acknowledgement}
%
%This work was supported by the University of Sharjah (Project \# 1602040634-P). 

\section*{References}

%\bibliography{biblio}


  %\newpage
  
%\section*{Author contributions statement}

%Both co-authors contributed to the development of the idea behind this work. A.A. designed and carried out the experimental measurements. A.A. and A.S.E.  derived the mathematical expressions, and discussed and analyzed the results. A.A. wrote the manuscript. Both co-authors reviewed the manuscript and provided feedback.

%Conceived and designed the experiments
% 		Performed the experiments
% 		Analyzed the data
% 		Contributed materials/analysis tools
% 		Wrote the paper

%\section*{Additional information}

%\section*{Competing interests}
% The authors declare no competing financial interest.
 
%\section*{Data Availability}
%
%The data that support the findings of this study are available from the corresponding author upon request.

 \end{document}